\documentclass{ws-procs961x669}            
\usepackage{amsmath}
\usepackage{amssymb}
\usepackage{hyperref}
\usepackage{subfigure}
\usepackage{bm}
\usepackage{aas_macros}
\usepackage{orcidlink}
\usepackage{physics}
\usepackage{mathtools,slashed}
\usepackage{relsize}

\hypersetup{
	colorlinks=true,
	linkcolor=blue,
	filecolor=magenta,
	urlcolor=red,
	citecolor=magenta,
	pdftitle={Sharelatex Example},
}

\begin{document}
\title{Natural evidence for fuzzy sphere noncommutative geometry: super-Chandrasekhar white dwarfs}

\author{Surajit Kalita\orcidlink{0000-0002-3818-6037}}
\address{Department of Physics, Indian Institute of Science, Bangalore-560012, India\\
E-mail: surajitk@iisc.ac.in}

\author{T. R. Govindarajan\orcidlink{0000-0002-8594-0194}}
\address{The Institute of Mathematical Sciences, Chennai-600113, India\\
	E-mail: trg@imsc.res.in}

\author{Banibrata Mukhopadhyay\orcidlink{0000-0002-3020-9513}}
\address{Department of Physics, Indian Institute of Science, Bangalore-560012, India\\
	E-mail: bm@iisc.ac.in}

\begin{abstract}
Noncommutative geometry is one of the quantum gravity theories, which various researchers have been using to describe different physical and astrophysical systems. However, so far, no direct observations can justify its existence, and this theory remains a hypothesis. On the other hand, over the past two decades, more than a dozen over-luminous type Ia supernovae have been observed, which indirectly predict that they originate from white dwarfs with super-Chandrasekhar masses $2.1-2.8 \rm\,M_\odot$. In this article, we discuss that considering white dwarfs as squashed fuzzy spheres, a class of noncommutative geometry, helps in accumulating more mass than the Chandrasekhar mass-limit. The length-scale beyond which the effect of noncommutativity becomes prominent is an emergent phenomenon, which depends only on the inter-electron separations in the white dwarf.
\end{abstract}

\keywords{noncommutative geometry, white dwarfs, type Ia supernova, mass-limit.}

\bodymatter

\section{Introduction} 
Noncommutative geometry, a theory of quantum gravity, has been used for many decades to explain physical systems. Riemann quoted in 1854 that ``{\em \dots it seems that empirical 	notions on which the metrical determinations of space are founded, the notion of a solid body and a ray of light cease to be valid for the infinitely small. We are therefore quite at liberty to suppose that the metric relations of space in the infinitely small do not conform to hypotheses of geometry; and we ought in fact to suppose it, if we can thereby obtain a simpler explanation of phenomena \dots}''. In 1930s, Bronstein, using Heisenberg uncertainty principle and the features of Einstein's general relativity (GR), first argued that gravitational dynamics does not allow to measure arbitrarily small spacetime distances, concluding that the notions of classical Riemmanian geometry should be duly modified. The formalism of a fuzzy sphere, first introduced in 1992 \cite{1992CQGra...9...69M}, was later modified to show equivalence between noncommutativity (NC) and Landau levels in the presence of magnetic fields \cite{2015JPhA...48C5401A}. Moreover, NC alters the spacetime metric \cite{2006LNP...698...97N}, which results in shifting the event horizon and removing the essential singularity at the center of a black hole \cite{2017EPJC...77..577K}. etc. Unfortunately, there is no direct way to confirm the existence of NC like most of the other quantum gravity formalism, which compels it to remain a hypothesis.

In astrophysics, a white dwarf (WD) is the end state of a star with mass $\lesssim (10\pm 2) \rm\,M_\odot$. Chandrasekhar showed that the maximum mass of a carbon-oxygen non-magnetized and non-rotating WD is about $1.4\rm\,M_\odot$ \cite{1931ApJ....74...81C}. Above this mass-limit, they burst out to produce type Ia supernovae (SNe\,Ia) with nearly similar luminosities. However, observations of various peculiar over-luminous SNe\,Ia, such as SN\,2003fg, SN\,2006gz, SN\,2007if, SN\,2009dc, and many more, suggested that they were originated from super-Chandrasekhar WDs with mass $2.1-2.8\rm\,M_\odot$ \cite{2006Natur.443..308H,2010ApJ...713.1073S}. Over the years, different models including rotation, magnetic fields, modified gravity, ungravity effects, generalized Heisenberg uncertainty principle, to name a few, have been proposed to explain these massive WDs. However, since these super-Chandrasekhar WDs have not been observed directly, no one can single out the correct theory so far.

This article shows that the squashed fuzzy sphere algebra can quantize the energy dispersion relation, which further alters the equation of state (EoS) of the degenerate electrons present in a WD, resulting in the increase in mass of the WD beyond the Chandrasekhar mass-limit. It also discusses that the length-scale below which NC is prominent depends on the inter-electron separation. Hence, the inference of super-Chandrasekhar WDs can indirectly predicts the existence of NC in high-density regimes.

\section{Squashed fuzzy sphere and equation of state}
The formalism of a fuzzy sphere is 
analogous to the angular momentum algebra in ordinary quantum mechanics, where the position coordinates mimic the angular momentum variables. In an $N$-dimensional irreducible representation of $\mathtt{SU}(2)$ group, the commutation relation for the coordinates of fuzzy sphere, $X_i$ (i=1,2,3), is given by \cite{1992CQGra...9...69M}
\begin{equation}\label{Eq: Fuzzy sphere}
\comm{X_i}{X_j} = i\frac{k \hbar}{r} \epsilon_{ijk} X_k,
\end{equation}
where $k = 2r^2/\hbar\sqrt{N^2-1}$, $\hbar = h/2\pi$, $h$ is the Planck constant and $r$ the radius of the fuzzy sphere. 

\begin{figure}[htpb]
	\centering
	\includegraphics[scale=0.5]{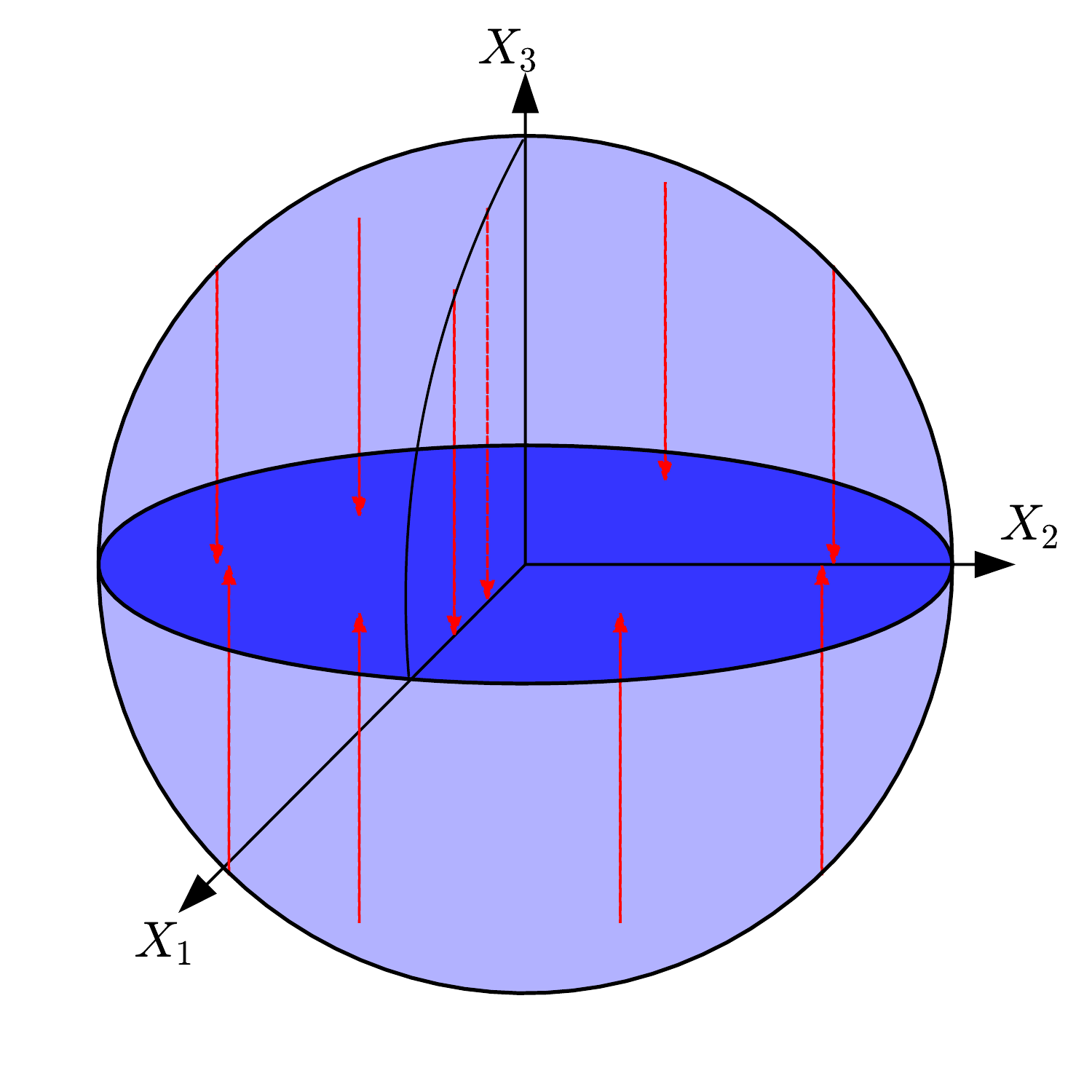}
	\caption{An illustrative diagram of a squashed fuzzy sphere which is obtained by projecting all the points of the fuzzy sphere on an equatorial plane.}
	\label{Fig: squashed fuzzy sphere}
\end{figure}
\begin{figure}[htp]
	\centering
	\includegraphics[scale=0.45]{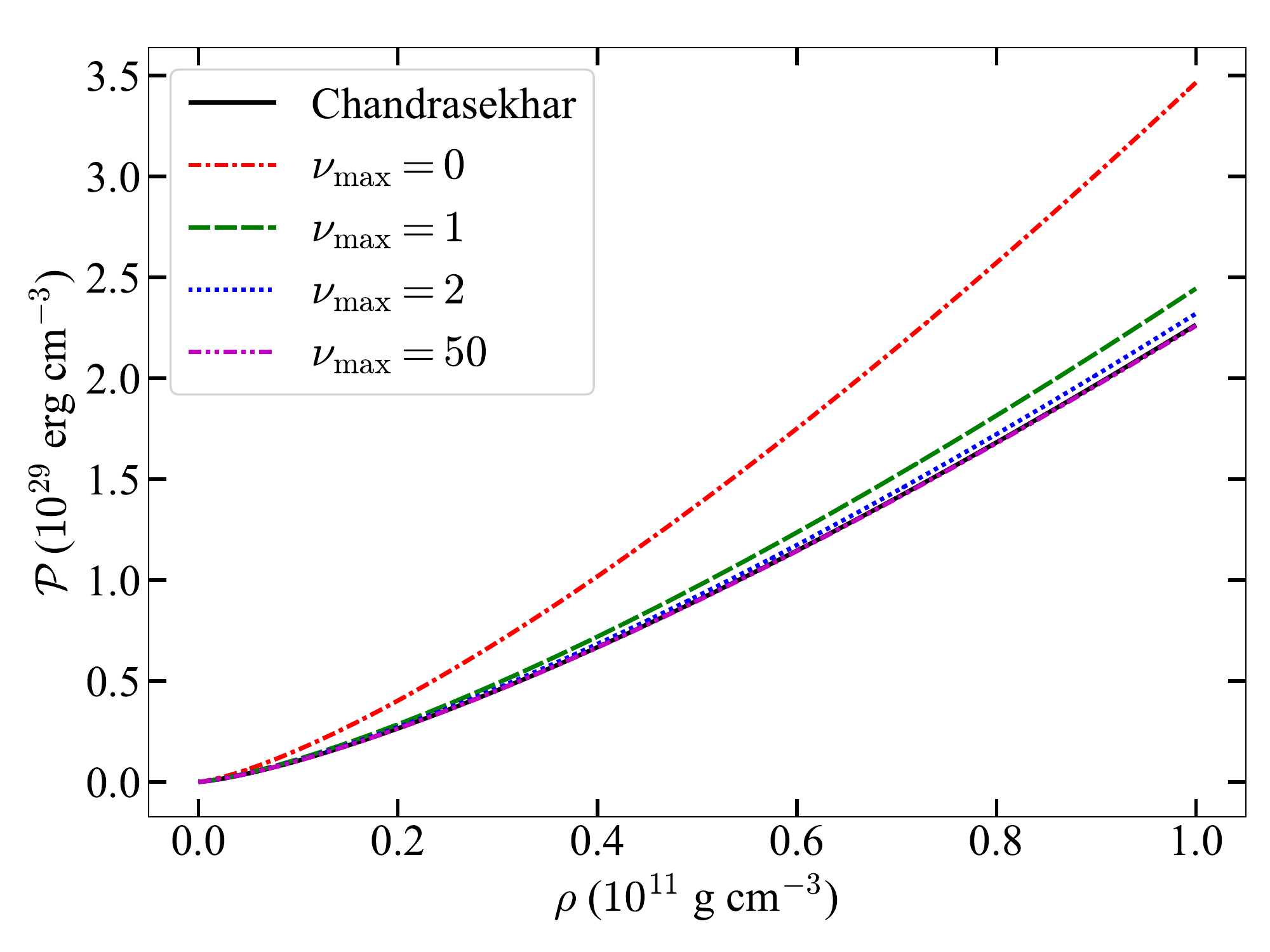}
	\caption{EoS for degenerate electrons in a squashed fuzzy sphere with various $\nu_\mathrm{max}$ along with the Chandrasekhar's EoS. $\nu_\mathrm{max}=0$ means electrons occupy only the ground energy level, $\nu_\mathrm{max}=1$ means they occupy both ground and first energy levels, and so on. Chandrasekhar's EoS corresponds to $\nu_\mathrm{max}\to\infty$.}
	\label{Fig: Fuzzy EoS}
\end{figure}

A squashed fuzzy sphere is a simple geometrical construction obtained from the fuzzy sphere by projecting all the points to an equatorial plane as shown in Figure~\ref{Fig: squashed fuzzy sphere}. The equatorial plane can be in any direction, and for illustration, we show the $X_1-X_2$ plane. Using the above relation, the commutation relation for this squashed fuzzy sphere is given by \cite{2015JPhA...48C5401A}
\begin{equation}\label{Eq: Squashed fuzzy sphere}
\comm{X_1}{X_2} = \pm i \frac{k \hbar}{r} \sqrt{r^2 - X_1^2 - X_2^2}.
\end{equation}
Note that NC vanishes at the surface of the sphere. The square of the quantized energies in a squashed fuzzy sphere, obtained from the Dirac operator, is given by \cite{2021arXiv210106272K}
\begin{align} \label{Eq: enrgy eigenvalues}
E_{l,m}^2 = \frac{2 \hbar c^2}{k \sqrt{N^2-1}} \left\{l(l+1)-m(m\pm 1)\right\},
\end{align}
where $c$ is the speed of light and $l$, $m$ are the quantum numbers with $l$ taking all the integer values from $0$ to $N-1$ and $m$ taking all the integer values from $-l$ to $l$. Moreover, the relations between the Cartesian and spherical polar coordinates are $X_1=r \sin\theta \cos\phi$ and $X_2=r \sin\theta \sin\phi$. Therefore, from Equation \eqref{Eq: Squashed fuzzy sphere}, the squashed fuzzy sphere algebra in spherical polar coordinates is given by
\begin{align}
\comm{\sin\theta \cos\phi}{\sin\theta \sin\phi} &= \pm i \frac{k \hbar}{r^2} \cos\theta.
\end{align}
A squashed fuzzy sphere is such that it actually provides a NC between its azimuthal and polar coordinates at the surface of the fuzzy sphere. This is because the squashed plane in a fuzzy sphere can be any of its equatorial planes and there is no particular direction for it, which means that the squashed fuzzy sphere has a rotational symmetry about the equatorial plane. There is no NC along the $r$-direction and an electron with mass $m_\mathrm{e}$ does not experience NC along the radial direction. Defining $p_r$ to be the momentum of the electron in $r$-direction, considering Equation~\eqref{Eq: enrgy eigenvalues}, we obtain the total energy dispersion relation in the squashed fuzzy sphere as \cite{2021arXiv210106272K}
\begin{align}\label{Eq: Fuzzy dispersion relation: v1}
E^2 = p_r^2 c^2 + m_\mathrm{e}^2 c^4 \left[1+ \{l(l+1)-m(m \pm 1)\} \frac{2 \hbar}{m_\mathrm{e}^2 c^2 k \sqrt{N^2-1}}\right].
\end{align}
In the large $N$ limit, this expression reduces to \cite{2021arXiv210106272K}
\begin{align}\label{Eq: Fuzzy dispersion relation}
E^2 = p_r^2 c^2 + m_\mathrm{e}^2 c^4 \left(1+ 2\nu \frac{2 \hbar}{m_\mathrm{e}^2 c^2 k}\right),\qquad \nu \in \mathbb{Z}^{0+}.
\end{align}
It is evident that $1/k$ behaves as the strength of NC. Note that a similar dispersion relation occurs in the case of a planar NC \cite{2021IJMPD..3050034K}. We can assume a series of concentric squashed fuzzy spheres, such that at the surface of each spheres, the above relation is valid. Nevertheless, $k\propto r^2$ and hence $1/k$ decreases from center to the surface. Now, applying simple statistical mechanics techniques, we obtain the EoS for the degenerate electrons present in a WD, given by \cite{2021arXiv210106272K,1991ApJ...383..745L}
\begin{align}\label{Eq: p_F rho relation}
\text{pressure\quad} \mathcal{P} &= \frac{2 \rho^{2/3}}{\xi h \mu_\mathrm{e}^{2/3} m_\mathrm{p}^{2/3}} \mathlarger{\mathlarger{\sum}}_{\nu=0}^{\nu_\mathrm{max}} g_\nu \left\{ p_\mathrm{F} E_\mathrm{F} - \left(m_\mathrm{e}^2 c^3 +2\nu \frac{2\hbar c}{k}\right) \ln(\frac{E_\mathrm{F} + p_\mathrm{F}c}{\sqrt{m_\mathrm{e}^2c^4 + 2\nu \frac{h c^2}{\pi k}}}) \right\},\\
\text{density\quad} \rho &= \frac{64 \mu_\mathrm{e} m_\mathrm{p} \left(2\nu_\mathrm{max}+1\right)^3p_\mathrm{F}^3}{\xi^3 h^3},
\end{align}
where $\xi= k h \left(\rho/\mu_\mathrm{e} m_\mathrm{p}\right)^3$, $m_\mathrm{p}$ is the mass of the proton, $\mu_\mathrm{e}$ is the mean molecular weight per electron, $g_\nu$ is the degeneracy factor, and $p_\mathrm{F}$ and $E_\mathrm{F}$ are respectively the Fermi momentum and Fermi energy of the electron gas. $\xi$ needs to be chosen appropriately such that at the low-density limit, EoS is same as the Chandrasekhar's EoS, where NC is almost negligible. Figure \ref{Fig: Fuzzy EoS} shows EoS for the degenerate electrons when they occupy different energy levels. It is evident that if the electrons occupy more levels, it tends to the Chandrasekhar's EoS and NC is not prominent in that density regime.

\section{Stellar structure model} 
\begin{figure}[htp]
	\centering
	\includegraphics[scale=0.45]{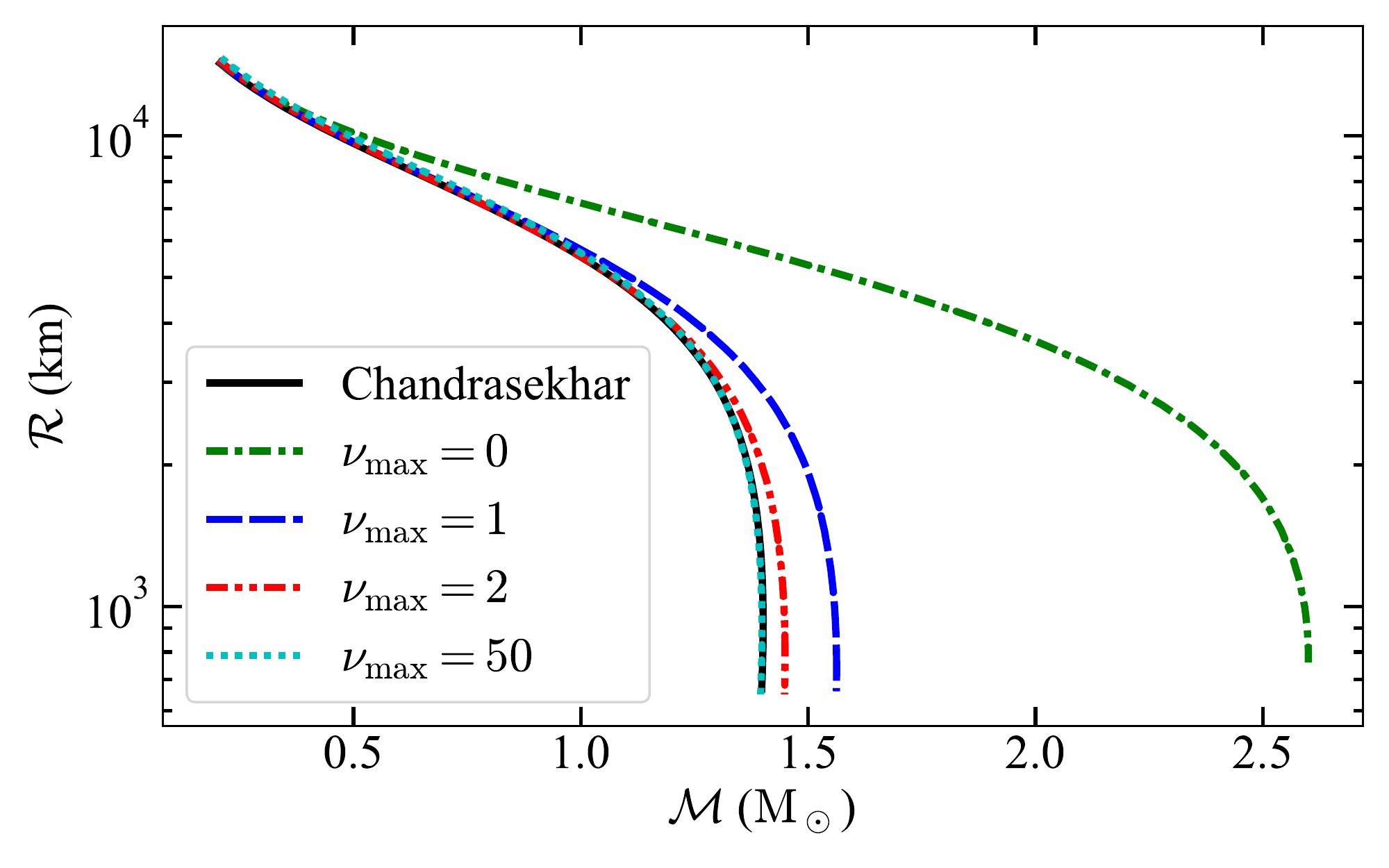}
	\caption{Mass--radius relations of the WDs in the presence of NC when various energy levels are occupied.}
	\label{Fig: Fuzzy M_R}
\end{figure}

To obtain the stellar structure of any compact object in GR, one needs to solve the Tolman-Oppenheimer-Volkoff or TOV equations along-with EoS of the constituent particles. The TOV equations for a non-rotating and non-magnetized star are given by \cite{2009igr..book.....R}
\begin{equation}\label{TOV}
\begin{aligned}
\dv{\mathcal{M}}{r} &= 4\pi r^2\rho,\\
\dv{\mathcal{P}}{r} &= -\frac{G}{r^2}\left(\rho+\frac{\mathcal{P}}{c^2} \right)\left(\mathcal{M}+\frac{4\pi r^3 \mathcal{P}}{c^2}\right) \left(1-\frac{2G\mathcal{M}}{c^2r}\right)^{-1},
\end{aligned}
\end{equation}
where $\mathcal{M}$ is the mass of the star inside a volume of radius $r$ and $G$ is Newton's gravitational constant. Note that NC affects the microscopic physics, while TOV equations describe the pressure and mass balances, which are macroscopic physics. Hence, in a semi-classical approach, TOV equations remain unchanged to that for the classical GR formalism. Figure \ref{Fig: Fuzzy M_R} shows the mass--radius relation for the WDs in the presence of NC for various $\nu_\mathrm{max}$ along with the Chandrasekhar's mass--radius relation, which corresponds to $\nu_\mathrm{max}\to\infty$. It is evident from the figure that a non-rotating and non-magnetized WD can have a limiting mass of about $2.6\rm\,M_\odot$ if all the electrons occupy the ground level. The mass-limit decreases when more energy levels are filled, and finally coincides with the Chandrasekhar's original mass--radius curve when many levels are occupied furnishing a limiting mass of about $1.4\rm\,M_\odot$.

\subsection{Length scale in noncommutative geometry}
It is understood from Equation \eqref{Eq: Fuzzy dispersion relation} that NC is prominent if $2\hbar\gtrsim m_\mathrm{e}^2c^2k$, which implies that the system length-scale $\mathcal{L}\lesssim\mathcal{L}_\mathrm{eff}=\lambda_\mathrm{e}/\sqrt{\pi\xi}$ with $\lambda_\mathrm{e}$ being the Compton wavelength of the electron \cite{2021arXiv210106272K}. In the case of degenerate electrons, $\mathcal{L}$ is the inter-electron separations. It is evident from this expression that for the prominence of NC, $\mathcal{L}$ does not need to be the Planck scale $\mathcal{L}_\text{P}$ as opposed to the general belief. Rather it depends upon the system's length-scale. Indeed Salecker and Wigner showed that the new uncertainty in length-scale for a system has to be $\delta \sim \left(\mathcal{L} \mathcal{L}_\text{P}^2\right)^{1/3}$, and one needs to consider $\delta$ as the quantum measurement of length \cite{1958PhRv..109..571S}. Hence, for a WD, one can expect to observe a significant NC effect depending on the inter-electron separation, even though it is far from the Planck scale. In extremely high densities with inter-electron distance much less than the Compton wavelength, one can expect certain features of NC becomes prominent. One such feature is the notion of statistics, which needs localization of particles, gets distorted and can affect physical systems, such as the mass-limit of the WD.

\section{Conclusions}
We have demonstrated that NC may have a prominent effect in a system whose length-scale is greater than $\mathcal{L}_\text{P}$. Whether NC has a significant effect or not depends upon the system's length-scale, and we show that it is an emergent phenomenon. We have used the formalism of a squashed fuzzy sphere where NC is at the surface of the sphere, i.e., between the azimuthal and polar coordinates. We have obtained the energy dispersion relation for this formalism and found that it is very similar to that for the Landau levels in the presence of magnetic fields. Hence, one may assume that NC mimics the internally generated magnetic field. It has a significant effect in a WD if the inter-electron separation is nearly less than one-tenth of the Compton wavelength of electrons. Such prominent NC may provide a super-Chandrasekhar limiting mass $2.6\rm\,M_\odot$ and it can explain the formation of peculiar over-luminous SNe\,Ia. In the future, if gravitational wave detectors detect these massive WDs directly, it can confirm the existence of NC more firmly \cite{2020ApJ...896...69K}.

\bibliographystyle{MG_proceedings}
\bibliography{bibliography}

\begin{thebibliography}{10}

\bibitem{1992CQGra...9...69M}
J.~{Madore}, {The fuzzy sphere}, {\em Classical and Quantum Gravity} {\bf 9},
  69 (Jan 1992).

\bibitem{2015JPhA...48C5401A}
S.~{Andronache} and H.~C. {Steinacker}, {The squashed fuzzy sphere, fuzzy
  strings and the Landau problem}, {\em Journal of Physics A Mathematical
  General} {\bf 48}, p. 295401 (Jul 2015).

\bibitem{2006LNP...698...97N}
V.~P. {Nair}, {\em {Noncommutative Mechanics, Landau Levels, Twistors, and
  Yang-Mills Amplitudes}}, in {\em Lecture Notes in Physics, Berlin Springer
  Verlag\/},  ed. S.~{Bellucci} 2006, p.~97.

\bibitem{2017EPJC...77..577K}
R.~{Kumar} and S.~G. {Ghosh}, {Accretion onto a noncommutative geometry
  inspired black hole}, {\em European Physical Journal C} {\bf 77}, p. 577 (Sep
  2017).

\bibitem{1931ApJ....74...81C}
S.~{Chandrasekhar}, {The Maximum Mass of Ideal White Dwarfs}, {\em \apj} {\bf
  74}, p.~81 (July 1931).

\bibitem{2006Natur.443..308H}
D.~A. {Howell}, M.~{Sullivan}, P.~E. {Nugent}, R.~S. {Ellis}, A.~J. {Conley},
  D.~{Le Borgne}, R.~G. {Carlberg}, J.~{Guy}, D.~{Balam}, S.~{Basa},
  D.~{Fouchez}, I.~M. {Hook}, E.~Y. {Hsiao}, J.~D. {Neill}, R.~{Pain}, K.~M.
  {Perrett} and C.~J. {Pritchet}, {The type Ia supernova SNLS-03D3bb from a
  super-Chandrasekhar-mass white dwarf star}, {\em \nat} {\bf 443}, 308
  (September 2006).

\bibitem{2010ApJ...713.1073S}
R.~A. {Scalzo}, G.~{Aldering}, P.~{Antilogus}, C.~{Aragon}, S.~{Bailey},
  C.~{Baltay}, S.~{Bongard}, C.~{Buton}, M.~{Childress}, N.~{Chotard},
  Y.~{Copin}, H.~K. {Fakhouri}, A.~{Gal-Yam}, E.~{Gangler}, S.~{Hoyer},
  M.~{Kasliwal}, S.~{Loken}, P.~{Nugent}, R.~{Pain}, E.~{P{\'e}contal},
  R.~{Pereira}, S.~{Perlmutter}, D.~{Rabinowitz}, A.~{Rau}, G.~{Rigaudier},
  K.~{Runge}, G.~{Smadja}, C.~{Tao}, R.~C. {Thomas}, B.~{Weaver} and C.~{Wu},
  {Nearby Supernova Factory Observations of SN 2007if: First Total Mass
  Measurement of a Super-Chandrasekhar-Mass Progenitor}, {\em \apj} {\bf 713},
  1073 (April 2010).

\bibitem{2021arXiv210106272K}
S.~{Kalita}, T.~R. {Govindarajan} and B.~{Mukhopadhyay}, {Super-Chandrasekhar
  limiting mass white dwarfs as emergent phenomena of noncommutative squashed
  fuzzy spheres}, {\em International Journal of Modern Physics D} , p. (in
  press) (November 2021).

\bibitem{2021IJMPD..3050034K}
S.~{Kalita}, B.~{Mukhopadhyay} and T.~R. {Govindarajan}, {Significantly
  super-Chandrasekhar mass-limit of white dwarfs in noncommutative geometry},
  {\em International Journal of Modern Physics D} {\bf 30}, p. 2150034 (April
  2021).

\bibitem{1991ApJ...383..745L}
D.~{Lai} and S.~L. {Shapiro}, {Cold Equation of State in a Strong Magnetic
  Field: Effects of Inverse beta -Decay}, {\em \apj} {\bf 383}, p. 745
  (December 1991).

\bibitem{2009igr..book.....R}
L.~{Ryder}, {\em {Introduction to General Relativity}} (Cambridge University
  Press, 2009).

\bibitem{1958PhRv..109..571S}
H.~{Salecker} and E.~P. {Wigner}, {Quantum Limitations of the Measurement of
  Space-Time Distances}, {\em Physical Review} {\bf 109}, 571 (Jan 1958).

\bibitem{2020ApJ...896...69K}
S.~{Kalita}, B.~{Mukhopadhyay}, T.~{Mondal} and T.~{Bulik}, {Timescales for
  Detection of Super-Chandrasekhar White Dwarfs by Gravitational-wave
  Astronomy}, {\em \apj} {\bf 896}, p.~69 (June 2020).

\end{thebibliography}

\end{document}